 \documentstyle[12pt]{article}
 \newcommand{\widebar}{\bar}
 \newcommand{\eq}{\begin{equation}}
 \newcommand{\ee}{\end{equation}}

 \newcommand{\s}{{\sigma}}

\begin{document}
 \vspace{0.2in}
 \begin{center}
 \large\bf
 Transitions Between Hall Plateaus and the Dimerization Transition of a
 Hubbard Chain \\
 \vspace{0.5in}
 \normalsize\rm
 Dung-Hai Lee$^{(a)}$ and Ziqiang Wang$^{(b)}$ \\
 \vspace{0.2in}
 \em
 $^{(a)}$Department of Physics, University of California at Berkeley, 
 Berkeley, CA 94720 \\
 \vspace{0.2in}
 $^{(b)}$Department of Physics, Boston University, Boston, MA 02215 \\

 \vspace{1.0in}
 \bf
 Abstract
 \end{center}
 \vspace{0.2in}
 \vbox{\narrower\smallskip
 \rm

 We show that the plateau transitions in the quantum Hall effect is the same
 as the dimerization transition of a half-filled, one dimensional, $U(2n)$ 
 Hubbard model at $n=0$. We address the properties of the latter  
 by a combination of perturbative renormalization group and Monte Carlo
 simulations. Results on both critical and off-critical properties
 are presented.
 
 \smallskip}
 \vskip 1cm
 \begin{verbatim}
 PACS numbers: 73.50.Jt, 05.30.-d, 74.20.-z
 \end{verbatim}
 \newpage
 \setlength{\baselineskip}{.375in}
 \rm
 
 The integer plateau transitions in the quantum Hall effect have
 attracted considerable interests recently \cite{review}.
 Aside from a recent speculation \cite{lwk}, the effect
 of electron-electron interaction on the
 transition remains largely unknown. Indeed, most of the recent theoretical 
 works on this subject are based on numerical
 analyses of models of {\it non-interacting} electrons \cite{review}.
 Nonetheless, they have generated a wealth of 
 interesting results. Among them, a consensus on the value of the 
 localization length exponent $\nu\approx 2.3$ has been reached
 \cite{review}. Remarkably this value,
 obtained without considering electronic interactions, agrees
 excellently with the experimental findings \cite{wei}.
 This agreement is even
 {\it surprising}  in view of the recently measured dynamical 
 exponent $z=1$ \cite{engel} instead of the non-interacting 
 value $2$.
 
 Even within the non-interacting theory, our understanding of the plateau
 transition is still limited. In this paper, we show that the 
 latter is equivalent to the zero temperature, dimerization transition of a 
 half-filled, $2n$-component Hubbard chain, with $n=0$.
 Thus from now on we shall view 
 physics from two different angles - one associated with the plateau 
 transition, and the other with the dimerization transition. For examples,
 the two neighboring plateau phases correspond to the two  
 dimer phases when the hopping integrals in the Hubbard model are
 allowed to alternate between neighboring bonds;
 Tuning the Fermi energy corresponds to adjusting the degree of 
 staggering in the hopping integrals; The 
 variance in the Aharonov-Bohm phases as electrons traverse the 
 edges of the randomly distributed quantum Hall droplets is proportional to
 the Hubbard $U$. 

 Crudely speaking, the behaviors of the half-filled, $n=0$ and $n=1$ 
 Hubbard chains are quite {\it similar}. For example, in both cases a 
 nonzero 
 single-particle gap exists for all value of $U/t$ (i.e. Hubbard 
 interaction/ the averaged hopping matrix elements). For $n=0$ this is the
 familiar statement that the disorder averaged 
 single-particle Green's function is always short-ranged. 
 When the hopping strength alternates between neighboring bonds
 (Fig.1b), both models exhibit nonzero spin gaps $\Delta_s$ \cite{loc}.
 The spin gap vanishes continuously as the translational symmetry in
 hopping is gradually restored.
 Specifically, if we define 
 $R\equiv (t_1-t_2)/(t_1+t_2)$, where $t_1$ and $t_2$ are values of the
 alternating hopping matrix elements shown in Fig.1b, 
 $\Delta_s\sim \mid R\mid^{\nu}$ as 
 $R\rightarrow 0$. 
 One of the differences between the $n=0$ and $n=1$ models shows up in 
 the value of $\nu$. For $n=1$, $\nu=2/3$ \cite{affleck1}, while for $n=0$,
 $\nu=2.33\pm.03$ (Fig.2, inset (a)) \cite{review}.
 Furthermore, at $R=0$ while the gapless spin 
 excitations in $n=1$ can be identified with those of the $n=1$ free 
 fermion chain \cite{affleck1}, we found that {\it it is not so} for $n=0$. 
 There are also non-critical properties that differentiate between
 the two cases. For example, the spontaneous staggered magnetization (SSM)
 is zero for all $R$ for $n=1$, whereas for $n=0$ it is always nonzero 
 (Fig.~3 inset(a))).  The latter appears rather counter-intuitive,
 since it requires the existence of a SSM 
 {\it in the presence of a spin gap} \cite{ms}!
 However, when stated in terms of the plateau transition it becomes
 very natural --- the density of states (DOS) stays nonzero across the 
 transition \cite{wegner1}.
 
 We have computed the spin-spin correlation 
 function (or the two-particle Green's function in the original language)
 at $R=0$ for the $n=0$ Hubbard chain using Monte Carlo method.  
 The results give for $x_1$, the scaling dimension of the spin, 
 $x_1\approx 0$ (Fig.3). 
 This value is consistent with a non-critical DOS
 across the plateau transition \cite{wegner1}.
 The calculation also determines the dimension, $x_2$,
 of the leading scaling operator formed by the operator product
 of two spins, being $x_2= -0.60\pm.02$ (Fig.3). 
 From $x_2$ we can deduce the fractal dimension 
 of the eigen-wavefunctions at $R=0$ \cite{wegner2} via 
 $D(2)=2+x_2=1.40\pm.02$.
 These values agree with those obtained in Refs.\cite{pook,huo}
 while differ somewhat from those of Refs.\cite{chaldan,hucksw}.
 In the literature there has been a misnomer, namely, taking $-x_2$ as the 
 decay-exponent of the spin-spin correlation function (or the two-particle
 Green's function).
 In Fig.~3, inset(b) we show
 the low temperature, equal-time spin-spin correlation 
 for a Hubbard chain of a {\it fixed} length $M$. One does see that it decays 
 with an exponent $-x_2$. Only after studying the finite size scaling with
 respect to $M$, is one able to conclude that this
 decay is the property of the scaling function ${\cal F}$ in Eq.(7).
 
 The Lyapunov exponents of the transfer matrix
 are often studied in {\it typical} long disordered cylindrical samples.
 In the Hubbard model language, it corresponds to
 Hubbard-Stratonovich (H-S) decouple the Hubbard interaction and study the
 single-particle excitation energies in a {\it specific} (but typical) 
 H-S field at a very low temperature. The Lyapunov exponents
 $\mu_k, \ k=1,2,3,...$ are defined as $\mu_k\equiv M\Delta_k$, where 
 $\Delta_k$ is the $k$th lowest single-particle excitation energy.
 For small $\mu_k$, we found $\mu_k=A+kB$
 where $A$ and $B$ depend only on 
 $U$ and $t$ 
 (Fig.~2 inset (b)). 
 This result is consistent with that the single-particle Greens 
 function is conformal invariant under a typical H-S field. 
 In addition, it suggests that the associated decay exponent, 
 $\eta_{\rm typical}$, is
 given by $\eta_{\rm typical}=2A/B=0.78\pm.02$ \cite{cardy}.
 In the literature one often uses $\eta_{\rm typical}=A/\pi$, and ignores the fact  
 that the spin-wave velocity can deviate from unity \cite{cc}.
 We have explicitly checked that while $A/\pi$ changes with $U/t$, 
 $2A/B$ does not. In a {\it fixed} H-S field,
 the decay exponent for the 
 spin-spin correlation function is twice that of
 the single-particle one, thus $x_{1,\rm typical}=\eta_{\rm typical}
 =0.78\pm.02$. 
 The departure of $x_{1,\rm typical}$ from $x_1$ usually signifies
 the existence of an infinite hierarchy of primary operators 
 $\{O_m\}$ formed by the product of $m$ $SU(2n)$ spin operators \cite{ludwig}.
 
 Now we step into the more technical part of the paper.
 We use the Chalker-Coddington network model to describe the plateau 
 transition \cite{lwk,cc}.
 Specifically, one considers a
 square lattice and draws arrows with a fixed chirality around half of the
 plaquettes (Fig.~1a) representing the 
 semiclassical electron orbits at the edges 
 of quantum Hall droplets.
 Quantum tunneling is only allowed at the vertices.
 Away from these vertices, the electrons propagate 
 along the directed links while accumulating Aharonov-Bohm phases.
 In a previous paper \cite{dhl}, one of us recognized that if one replaces 
 these link Aharonov-Bohm phases by random {\it tunneling phases}, the 
 resulting network model is equivalent to an $SU(2n)\mid_{n\rightarrow 0}$ 
 quantum spin chain. We show below that the original
 network model of Ref.\cite{cc} is in fact a $U(2n)\mid_{n\rightarrow 0}$
 Hubbard chain. 

 A faithful representation of the network model is depicted in Fig.~1b, where
 electrons move along the zigzag paths and tunnel
 across the lines joining them. By adjusting the tunneling matrix
 elements, one can 
 reproduce the original 2x2 scattering matrix at each tunneling vertex. 
 From now on we take each zigzag path as a constant-$x$ curve and 
 let $y$ measure the displacement along it. We choose the coordinates
 so that the tunneling vertices are at $(x,y)={\rm integers}$. 
 The (non-interacting) Hamiltonian for the network model is then
 \eq
 {H}=\sum_x (-1)^{x} \int dy 
 \psi^\dagger (x,y)\left[{{\partial_y}\over i}-w(x,y)\right]\psi (x,y)
 +\sum_{x,y} t_{x,y}^\prime\left[\psi^\dagger (x+1,y)\psi (x,y)
 + h.c\right].
 \ee
 Here $\psi$ is the electron annihilation operator, 
 $t_{x,y}^\prime$ are the tunneling matrix elements, and   
 $w$ is a local random variable satisfying  
 $<w(x,y)w(x',y')>=U'\delta_{x,x'}\delta(y-y')$.
 Upon accumulation, the latter gives rise to the Aharonov-Bohm phases. 
 As in transport theories, we write down the generating functional 
 ${\cal Z}[w]=\int D[\widebar{\psi},\psi] exp\{-{\cal S}\}$ 
 for the Green's function at the Fermi energy ($E=0$), where
 \begin{eqnarray}    
 {\cal S}&=&\sum_{x,p}\int\! dy (ih) S_p
 \widebar{\psi}_p(x,y)\psi_p(x,y) - H\mid_{\psi\rightarrow\psi,\psi^\dagger
 \rightarrow\widebar{\psi}}.
 \end{eqnarray}           
 In the above, the last term
 stands for replacing
 the field operators by Grassmann variables in Eq.~(1),
 $p=+(-)$ generates the advanced (retarded) Green's function,
 $S_p\equiv {\rm sign}(p)$, and $h$ is a positive infinitesimal.
 Our strategy is to make transformations that turn Eq.~(2) into
 a 1+1 dimensional Euclidean action of an interacting
 quantum theory in one space dimension \cite{fradkin}.
 First, we let $U'=U\epsilon$ and $t'=t\epsilon$ where $\epsilon
 \rightarrow 0$. (We have checked that doing so neither 
 affects our ability to tune through the transition, nor does it change the
 universality class.)
 By regarding $\epsilon$ as the lattice spacing 
 in $y$-direction, we are able to write down an action that is
 continuous in $y$. 
 Next, we let $\psi_p\rightarrow\psi_p(i\psi_p)$ and $\widebar{\psi}_p\rightarrow
 -i \widebar{\psi}_p(\bar{\psi}_p)$ for even (odd) $x$'s, and integrate
 out $w$ using the replica trick to obtain
 \begin{eqnarray}
 {\cal S}_R&=&\int\! dy\sum_{x,a}\left[\widebar{\psi}_a(x,y)
 \partial_y\psi_a(x,y)+(-1)^xh S_p
 \widebar{\psi}_a(x,y)\psi_a(x,y) 
 +{{U}\over 2}\widebar{\psi}_a(x,y)
 \psi_a(x,y)\right] \\  
 &+&{{U}\over 2}\int\! dy\sum_{x,a\ne b}
 \widebar{\psi}_a(x,y)\psi_a(x,y)
 \widebar{\psi}_b(x,y)\psi_b(x,y)  
 +\sum_{x,y,a}t_{x,y}\left[\widebar{\psi}_a(x+1,y)\psi_a(x,y)
 + h.c.\right]. \nonumber
 \end{eqnarray}
 Here $a,b=(p\alpha),(p'\beta)$ with replica indices $\alpha,\beta
 =1,...,n$. 
 In the following we shall take $t_{x,y}$ to be $y$-independent and satisfying
 $t_x=t_{x+2}$ \cite{note}.
 To tune through the plateau transition we adjust
 $R\equiv (t_1-t_2)/(t_1+t_2)$.
 Now if we regard $y$ as the Euclidean time $\tau$, 
 then ${\cal S}_R$ can be viewed as the 
 1+1 dimensional action of a quantum theory with the following Hamiltonian
 \begin{eqnarray}       
 H=\sum_{x,a}\left\{-h(-1)^xS_p\psi^\dagger_a(x)\psi_a(x)
 -t_x\left[\psi^\dagger_a(x+1)\psi_a(x)
 + h.c.\right]\right\}+{U\over 2}\sum_{x}[\sum_a\psi^\dagger_a(x)\psi_a(x)]^2 . 
 \end{eqnarray}           
 Eq.~(4) is the $U(2n)$ Hubbard Hamiltonian.
 Note that due to the transformations on $\psi$ and 
 $\widebar{\psi}$ the Green's function of Eq.(2) is related to that of Eq.(4)
 via $G_p(x,x';y,y')\rightarrow \theta(x,x')G_a(x,x';\tau,\tau')$,  
 where $\theta=-i(-1)^x$ if $x,x'$ are on the same sublattice and 
 $\theta=1$ otherwise. As a result,
 $G_p(x,x';y,y')G_{p'}(x',x;y',y)\rightarrow -(-1)^{x-x'}
 G_a(x,x';\tau,\tau')G_b(x',x;\tau',\tau)$.
 The positive infinitesimal $h$ in Eq.(2)
 becomes a small staggered magnetic field in the Hubbard
 model. Hence, the DOS corresponds to the SSM.


For $n=0$ Eq.(4) is particle-hole symmetric,
thus the system is half-filled. At $R=0$ there are
gapless spin excitations.
Since the latter in the half-filled
$n=1$ Hubbard model are known to be the same as those of the $U(2)$ 
free fermion model \cite{affleck1},
we first check whether the same is true for $n=0$. Thus
we study the stability of the free fermion spin sector against
a small Hubbard $U$. As usual
we linearize the free-fermion 
dispersion within an energy window around the left(L) and right(R) 
 Fermi points. Then we project the Hubbard interaction into this energy 
 window. That amounts to keeping four scattering amplitudes:
 $g_1$ (L-R backward), $g_2$ (L-R forward),
 $g_3$ (umklapp), and $g_4$ (L-L or R-R forward) \cite{note7}.
 We then perform a one-loop renormalization group calculation and obtain
 the following flow equations:
 \eq
 {{dg_1}\over {dl}}=-{1\over {2\pi}}\left[2ng^2_1+2(n-1)g^2_3\right], 
 \ 
 {{dg_2}\over {dl}}={1\over {2\pi}}\left[g^2_3-g^2_1\right], \ 
 {{dg_3}\over {dl}}=-{1\over {\pi}}\left[(2n-1)g_1-2g_2\right]g_3,
 \ee
 and as usual, $dg_4/dl=0$. 
 In the above we have set $t=1$ and the bare $g$-values are
 $g_i=U, \ i=1,\dots,4$.
 Eq.~(5) shows that $n=1$ is quite special. For
 $n\ge 1$ $g_1\rightarrow 0$ upon renormalization while $g_2$ and $g_3$ flow
 to large values \cite{marston}.
 (At $n=1$, $g_1\rightarrow 0$ allows the solution of the spin 
  sector of the Hubbard model in term of that of the free fermions.)
  However, for $n<1$, all $g_{1,2,3}$ flow to
  large values upon renormalization. 
  Therefore, the plateau transition is governed by a strong coupling
  fixed point of the Hubbard model at $n=0$.

 
 %
 
 To pin down the strong-coupling fixed point Hamiltonian at $R=0$,
 we have followed a recent method used by Nagaosa and 
 Oshikawa\cite{no} and showed that the half-filled 
 $U(2n)$ Hubbard model is equivalent to a $U(2n)/U(n)\times U(n)$ non-linear 
 $\s$-model \cite{pruisken}, which in turn is equivalent to an $SU(2n)$ 
 spin chain \cite{dhl,affleck2}. To support that conclusion, we have compared the 
 critical behaviors of the {\it typical} $\Delta_s$ for the $U(2n)$ Hubbard and 
 $SU(2n)$ spin chains at $n=0$. Specifically, we calculated the typical $\Delta_s$  
 for both models in long cylindrical systems 
 with linear dimension $M$ and time dimension $L$ ($L>>M$) \cite{unit}
 using the transfer matrix method.
 The results are given in Fig.~2, where the thermodynamic
 $\Delta_{s,\infty}$ is determined such that the data of $(M\Delta_{s,M})$
 for different $M$ fall onto a single scaling curve.
 In presenting the results we have  
 rescaled the $\Delta_s$ values at $R=0$, which amounts to
 adjusting a non-universal spin-wave velocity. 
 The excellent agreement between the scaling curves lends support 
 to the statement that these two models are indeed asymptotically equivalent.

 In order to study the strong coupling fixed point at $R=0$,
 we extend the quantum Monte Carlo algorithm
 \cite{monte} to general $n$.
 Specifically, we decouple the Hubbard interaction in Eq.~(4)
 via a spin-symmetric H-S transformation and trace out the Grassmann
 variables. The resulting partition function 
 $Z=\int\!{\cal D}[w]{\rm det}^{2n}P[w]$ is used to evaluate observables.
 The single-particle Green's function $G(xt,x't')\equiv 
 \langle\widebar{\psi}(x,t)\psi(x',t')\rangle$ is given by,
 \eq
 G(xt,x't')=Z^{-1}\int\!{\cal D}[w] G_w(xt,x't')
 {\rm det}^{2n}[w]P[w],
 \ee
 where $G_w$ denotes the single 
 particle Green's function computed in a 
 H-S field $\{w_i(\tau)\}$, ${\rm det}[w]$ is the fermion determinant
 and $P[w]= exp\{-\int\!
 d\tau\sum_i w_i(\tau)^2/2U\}/\cal{N}$ ($\cal{N}$ is a normalization
 factor). Note that as $n\rightarrow 0$ the 
 annealed average in Eq.~(6) becomes a quenched one. 
 Since in one dimension we expect the Hubbard model to have Lorentz 
 invariance, we approach the zero temperature and infinite size limit
 by finite size scaling of a sequence of systems with equal dimensionless
 spatial and temporal size $M$ \cite{unit}, and $h=\gamma/M^2$, with 
 $\gamma\ll1$. This choice of $h$ will be justified below. 

 An important and often raised issue is whether the network model
 is able to predict a nonzero DOS across the plateau transition.
 In terms of the Hubbard model, the issue concerns the existence of a
 SSM. 
 The staggered magnetization is obtained from the Monte Carlo results of
 the single-particle Green's function in Eq.~(6).
 The results are shown in Fig.~3, inset(a) for $R=0$ and $1$ as a function of $M$
 at $U/t=4$. They clearly suggest a nonzero DOS across the plateau transition.

 %
 
 To shed more light on the properties of the $n=0$ Hubbard model at $R=0$,
 we have computed the transverse spin-spin correlation function
 $\Gamma(x)\equiv\langle
 S_{+,\alpha}^{-,\beta}(x,t)S_{-,\beta}^{+,\alpha}(0,t)\rangle$ in 
 finite systems. 
 For  $1<<\mid x\mid<<M$, one can write down the scaling ansatz for $\Gamma$
 with the help of operator product expansion \cite{cardy},
 \eq
 \vert\Gamma(x,h,M)\vert\sim \left({1\over {\vert x\vert}}\right)^{2x_1}
 \left({{\vert x\vert}\over M}\right)^{x_2}
 {\cal F}(hM^2),
 \ee
 where the exponents $x_1$ and $x_2$ were defined earlier. 
 That $h$ scales with $M^{-2}$ in Eq.~(7) follows from
 the anticipation that the DOS is non-critical.
 The fact that data collapsing is achieved in Fig.~3 confirms the latter.
 The results are
 consistent Eq.~(7) with $x_1\approx 0$ and  $x_2= -0.60\pm.02$. 
 While the value of $x_1$ offers a consistency check on the notion
 of a non-critical DOS, that of $x_2$ determines the eigen-function fractal
 dimension at the transition \cite{review} through 
 $D(2)=2+x_2=1.40\pm.02$ \cite{pook,chalker}. 
 Note that in the thermodynamic limit $h\Gamma(0)$ corresponds to the
 ensemble averaged inverse participation ratio $P^{(2)}(E=0)$ introduced
 by Wegner \cite{wegner2}. As a check we have also determined 
 $D(2)$ from the scaling behavior 
 $h\Gamma(0)\propto M^{-D(2)}{\cal G}(hM^2)$. 
 The result is $D(2)=1.43\pm.04$.     
%
%

\noindent{Acknowledgment:} We thank J. Chalker, J. Gan, J.~B. Marston,
G. Murthy, R. Scalettar, S. Trugman, S. White for helpful discussions. 
DHL acknowledges the  
support from the LDRD program under DOE contract No. DE-AC03-76SF00098.

 \newpage
 \bibliographystyle{unsrt}

\begin{thebibliography}{99}
\bibitem{loc} For the $n=0$ case the spin gap is proportional to 
 the inverse localization length.
\bibitem{note}
  It has been checked that this choice is sufficient to
  study the plateau transition \cite{lwk,cc}.
\bibitem{unit}
  Here we have chosen a new temporal unit
  so that at $R=0$ the spin-wave velocity is of order unity.
\end{thebibliography}

  \newpage
  \centerline{FIGURE CAPTIONS}

 \ \

\noindent{\bf Fig. 1}. The network model (a) and its space(x)-time($\tau$)
representation (b). In (a) plaquettes marked with ``1'' are occupied by Hall 
droplets. Also shown in (b) is the corresponding Hubbard chain.

\ \

\noindent{\bf Fig. 2}. The scaling plots of the typical spin gap for the 
$n=0$ Hubbard ($U/t=4$) and spin chains ($M=16\to256$). Inset: (a) 
The exponent $\nu$. (b) The spectrum of Lyapunov exponents.

\ \

\noindent {\bf Fig. 3}. The scaling plot of $\Gamma$ in Eq.(7)
achieved with $x_1=0$. $M=192(\bigcirc)$, $160(\Box)$, $128(\Diamond)$,
$94(\bigtriangleup)$, $64(\bigtriangledown)$, $48(+)$, $32(\times)$.
The solid line gives a slope corresponding to $x_2=-0.60\pm.02$. 
Inset: (a) The size-dependence of the SSM in the Hubbard chain at 
critical point (circles) and for independent dimers (squares).
(b) $\ln\mid\Gamma(x)\mid$ for a $M=128$ chain at a temperature $\beta t=100$.
%
%
 \vspace*{\fill}
 \end{document}